\documentclass{ifacconf}

\usepackage{graphicx}      
\usepackage{natbib}        
\usepackage{xcolor}
\usepackage{amsmath, amsfonts} 
\usepackage{amssymb}
\begin{document}
\begin{frontmatter}

\title{Unbiased Parameter Estimation via DREM with Annihilators} 

\author[First]{Anton Glushchenko} 
\author[First]{Konstantin Lastochkin} 

\address[First]{V.A. Trapeznikov Institute of Control Sciences of RAS, Moscow, Russia (e-mail: aiglush@ipu.ru, lastconst@ipu.ru).}

\begin{abstract}                
In adaptive control theory, the dynamic regressor extension and mixing (DREM) procedure has become widespread as it allows one to describe {major} of adaptive control problems in unified terms of the parameter estimation problem of a regression equation with a scalar regressor. {However, when the system/parameterization is affected by perturbations, the estimation laws, which are designed on the basis of such equation, asymptotically provides only biased estimates.} In this paper, based on the bias-eliminated least-squares (BELS) approach, a modification of DREM procedure is proposed to annihilate perturbations asymptotically and, consequently, asymptotically obtain unbiased estimates. The theoretical results are supported with mathematical modelling and can be used to design adaptive observers and control systems.
\end{abstract}


\end{frontmatter}

\section{Introduction}
Using various {parameterizations}, the problems of controller and observer design for systems with {\it a priori} unknown parameters can be reduced to the one of \emph{online} identification of the regression equation parameters:
\begin{equation}\label{eq1}
z\left( t \right) = {\varphi ^{\top}}\left( t \right)\theta  + w\left( t \right){\rm{,}}   
\end{equation}
where $z\left( t \right) \in \mathbb{R}$ and $\varphi \left( t \right) \in {\mathbb{R}^n}$ are measurable for all $t \ge {t_0}$ regressand and regressor, $\theta  \in {\mathbb{R}^n}$ stands for unknown parameters, $w\left( t \right) \in \mathbb{R}$ denotes a bounded perturbation. 

For example, in \cite{b2} the problem of adaptive output-feedback control of a linear time-invariant nonminimum-phase dynamic system is reduced to the one of the parameter identification, and in \cite{b1} the problem of state observer design for the same class of systems is also transformed into the same problem. {A lot of approaches have been developed to deal with unknown parameters estimation task for equation \eqref{eq1} both in discrete (\cite{b210}) and continuous (\cite{b7}) time. In recent years, one of the most popular approaches to solve it is a dynamic regressor extension and mixing procedure (DREM) \cite{b3}, which reduces the regression equation \eqref{eq1} to a set of scalar regression equations and improves quality of unknown parameters estimates in perturbation-free case. In this note we are interested in improvement of such procedure properties in the presence of disturbance.}

DREM consists of a dynamic extension step ($l > 0$ is a filter parameter):
\renewcommand{\theequation}{2a}
\begin{equation}\label{eq2a}
\begin{array}{l}
\dot Y\left( t \right) =  - lY\left( t \right) + \varphi \left( t \right){z}\left( t \right){\rm{,\;}}Y\left( {{t_0}} \right) = {0_n}{\rm{,}}\\
\dot \Phi \left( t \right) =  - l\Phi \left( t \right) + \varphi \left( t \right){\varphi ^{\top}}\left( t \right){\rm{,\;}}\Phi \left( {{t_0}} \right) = {0_{n \times n}}{\rm{,}}
\end{array}   
\end{equation}
and mixing step
\renewcommand{\theequation}{2b}
\begin{equation}\label{eq2b}
{\cal Y}\left( t \right) = {\rm{adj}}\left\{ {\Phi \left( t \right)} \right\}Y\left( t \right).    
\end{equation}

Together, equations \eqref{eq2a} and \eqref{eq2b} allow one to transform the regression equation \eqref{eq1} into a set of scalar ones:
\renewcommand{\theequation}{\arabic{equation}}
\setcounter{equation}{2}
\begin{equation}\label{eq3}
{{\cal Y}_i}\left( t \right) = \Delta \left( t \right){\theta _i} + {{\cal W}_i}\left( t \right){\rm{,}}    
\end{equation}
where
\begin{displaymath}
\begin{array}{c}
{\cal Y}\left( t \right){\rm{:}} = {\rm{adj}}\left\{ {\Phi \left( t \right)} \right\}Y\left( t \right){\rm{,\;}}\Delta \left( t \right){\rm{:}} = {\rm{det}}\left\{ {\Phi \left( t \right)} \right\}{\rm{,}}\\
{\cal W}\left( t \right){\rm{:}} = {\rm{adj}}\left\{ {\Phi \left( t \right)} \right\}W\left( t \right){\rm{,}}\\
{\cal Y}\left( t \right) = {{\begin{bmatrix}
{{{\cal Y}_1}\left( t \right)}& \ldots &{{{\cal Y}_{i - 1}}\left( t \right)}&{\begin{array}{*{20}{c}}
 \ldots &{{{\cal Y}_n}\left( t \right)}
\end{array}}
\end{bmatrix}}^{\top}}{\rm{,}}\\
{\cal W}\left( t \right) = {{\begin{bmatrix}
{{{\cal W}_1}\left( t \right)}& \ldots &{{{\cal W}_{i - 1}}\left( t \right)}&{\begin{array}{*{20}{c}}
 \ldots &{{{\cal W}_n}\left( t \right)}
\end{array}}
\end{bmatrix}}^{\top}}{\rm{,}}\\
\dot W\left( t \right) =  - lW\left( t \right) + \varphi \left( t \right)w\left( t \right){\rm{,\;}}W\left( {{t_0}} \right) = {0_n}.
\end{array}    
\end{displaymath}

Based on the obtained system \eqref{eq3}, each $i^{\rm th}$ unknown parameter can be estimated independently using various identification laws. The degree of freedom for DREM is the choice of a method to extend the regressor \eqref{eq2a}. Instead of \eqref{eq2a}, the algorithms by \cite{b4,b5,b6} can also be used:
\renewcommand{\theequation}{4a}
\begin{equation}\label{eq4a}
\begin{array}{l}
\dot Y\left( t \right) = {\textstyle{1 \over T}}\left[ {\varphi \left( t \right)z\left( t \right) - \varphi \left( {t - T} \right)z\left( {t - T} \right)} \right]{\rm{,\;}}\\
\dot \Phi \left( t \right) = {\textstyle{1 \over T}}\left[ {\varphi \left( t \right){\varphi ^{\top}}\left( t \right) - \varphi \left( {t - T} \right){\varphi ^{\top}}\left( {t - T} \right)} \right]{\rm{,\;}}
\end{array}
\end{equation}
\renewcommand{\theequation}{4b}
\begin{equation}\label{eq4b}
 Y\left( t \right) = {\begin{bmatrix}
{{{\cal H}_1}\left( s \right)\left[z\left( t \right)\right]}\\
{{{\cal H}_2}\left( s \right)\left[z\left( t \right)\right]}\\
 \vdots \\
{{{\cal H}_n}\left( s \right)\left[z\left( t \right)\right]}
\end{bmatrix}} {\rm{,\;}}\Phi \left( t \right) = {\begin{bmatrix}
{{{\cal H}_1}\left( s \right)\left[\varphi \left( t \right)\right]}\\
{{{\cal H}_2}\left( s \right)\left[\varphi \left( t \right)\right]}\\
 \vdots \\
{{{\cal H}_n}\left( s \right)\left[\varphi \left( t \right)\right]}
\end{bmatrix}} {\rm{,}}   
\end{equation}
\renewcommand{\theequation}{4c}
\begin{equation}\label{eq4c}
\begin{array}{l}
\dot Y\left( t \right) =  - \Gamma \varphi \left( t \right){\varphi ^{\top}}\left( t \right)Y\left( t \right) + \Gamma \varphi \left( t \right)z\left( t \right){\rm{,\;}}\\
\Phi \left( t \right) = {I_n} - \Sigma \left( t \right){\rm{,}}\\
\dot \Sigma \left( t \right) =  - \Gamma \varphi \left( t \right){\varphi ^{\top}}\left( t \right)\Sigma \left( t \right){\rm{,\;}}\Sigma\left( {{t_0}} \right) = {I_n}{\rm{,}}
\end{array}    
\end{equation}
where $T > 0$ is a filtering window length, $\Gamma  = {\Gamma ^{\top}} > 0$ is an adaptive gain, ${{\cal H}_i}\left( s \right)\left[.\right]$ with $s : = {{{d}}\over{{dt}}}$ is an asymptotically stable linear filter ({\it e.g.}, ${{\cal H}_i}\left( s \right)\left[.\right]  = {\textstyle{1 \over {s + {\alpha _i}}}}\left[.\right]$, ${\alpha _i} > 0$).

In \cite{b3, b6, b7}, it is demonstrated that, when $w\left( t \right) \equiv 0$, the gradient-based identification law designed on the basis of equation \eqref{eq3} has a relaxed parametric convergence condition and improved transient quality compared to the gradient or least squares based laws designed using equation \eqref{eq1}. In \cite{b3,b6,b7,b8,b9,b10,b11,b12,b13}, various implementations of DREM procedure have been proposed, which differ from each other mainly by the filters ({\it e.g.}, \eqref{eq2a}, \eqref{eq4a}, \eqref{eq4b} or \eqref{eq4c}, etc.) for extension and/or the identification algorithms for $\theta_i$. 

The {chosen} extension scheme defines the properties of the regressor $\Delta\left(t\right)$ and the perturbations ${\mathcal{W}}_{i}\left(t\right)$. For example, scheme \eqref{eq4c} strictly relaxes the regressor persistent excitation condition, which is required to ensure exponential convergence of the unknown parameter estimates (\cite{b6}) in perturbation-free case. {A lot of studies investigated the influence of extension scheme on the estimates convergence for perturbed regressions.} In \cite{b16}, it is proved that the gradient-based identification law derived using equation \eqref{eq3} {with any extension scheme} ensures asymptotical convergence to the unknown parameters if \linebreak $\Delta  \notin {L_2}$ and $\Delta{{\cal W}_i} \in {L_1}$. In \cite{b11,b12}, { boundedness of $\tilde{\theta}\left(t\right)$ was shown for ${{\cal W}_i}\in L_{2}$}, and various estimation laws with finite time convergence and improved accuracy are developed for perturbed regressions. In \cite{b17}, an identification law is proposed which, in contrast to \cite{b3,b6,b7,b8,b9,b10,b11,b12,b13}, {using arbitrary extension scheme},  ensures asymptotic identification of the unknown parameters when the averaging conditions (${{\cal W}_i} \in {L_\infty }$ and $\int\limits_{{t_0}}^t {{\Delta ^{ - 1}}\left( s \right){{\cal W}_i}\left( s \right)ds < \infty } $) are met. In \cite{b18}, different discrete laws, which provide improved accuracy of unknown parameter estimation in case of perturbations, are compared. In \cite{b19,b20}, a new nonlinear filter is proposed, which ensures an arbitrary reduction of the steady-state parametric error under a certain regressor/perturbation ratio and independence of the regressor from the perturbation. In \cite{b21}, based on the method of instrumental variables, a {new extension scheme} is developed to guarantee asymptotic convergence of parameteric error for linear perturbed systems.

The main and general drawback of \cite{b3,b6,b7,b8,b9,b10,b11,b12,b13,b16,b17,b18,b19,b20,b21} is that the parametric convergence conditions are formalized in terms of properties of the perturbation ${\cal W}\left( t \right)$ from \eqref{eq3}. However, these conditions may never be met due to the features of the mixing procedure and the dynamic operators used at the extension step. In this case, only biased estimates can be obtained asymptotically using {scalar} regression equations \eqref{eq3}.

{Bias-eliminated least-squares
(BELS) \cite{b22,b23} is an approach to deal with {\it offline} discrete-time parameter estimation task for closed loop linear systems in the presence of coloured perturbations. In this case standard least squares identifier provides only biased estimates (\cite{b210}). Roughly speaking, when some controller structure is chosen and excitation conditions are met, BELS allows to compute and annihilate such bias from estimates.} In this study, based on the main idea of BELS approach, a modified DREM procedure is proposed, which in comparison with existing DREM based estimators \cite{b3,b6,b7,b8,b9,b10,b11,b12,b13,b16,b17,b18,b19,b20,b21} ensures that \emph{i}) the obtained estimates converge to arbitrarily {small} neighborhood of ideal parameters if {sufficiently large number of elements} of the regressor $\varphi\left(t\right)$ are independent from the perturbation $w\left(t\right)$,
\emph{ii}) the main conditions of convergence are formalized in terms of the perturbation and regressor of the original regression \eqref{eq1}. {Main contribution and distinctive feature of such design is to combine main ideas of DREM and BELS together for achievement of \emph{online} continuous-time asymptotically unbiased estimation in the presence of disturbance.}

\renewcommand{\theequation}{\arabic{equation}}
\setcounter{equation}{4}

\section{Problem Statement}
The aim is to design for \eqref{eq1} an \emph{online} estimation law, which, using measurable signals $\varphi \left( \tau  \right){\rm{,\;}}z\left( \tau  \right){\rm{\;}}{t_0} \le \tau  \le t$, ensures that the following conditions hold:
\begin{equation}\label{eq5}
\mathop {{\rm{lim}}}\limits_{t \to \infty } \left\| {\tilde \theta \left( t \right)} \right\| \le \varepsilon \left( T \right){\rm{,\;}}\mathop {{\rm{lim}}}\limits_{T \to \infty } \varepsilon \left( T \right) = 0{\rm{,}}
\end{equation}
where $T > 0$ is some parameter of the identification algorithm, and $\varepsilon {\rm{:\;}}{\mathbb{R}_ + } \mapsto {\mathbb{R}_ + }$.

\section{Main result}
In this section, a modified version of DREM procedure (\cite{b3,b7}) is designed on the basis of BELS approach (\cite{b22,b23}) previously applied for the discrete-time and offline identification. {Unlike \cite{b22, b23}, we solve not a linear system identification problem, but general perturbed linear regression equation estimation problem.}

To introduce the proposed estimator, we make some simple transformations of the linear regression equation \eqref{eq1}.
First of all, the linear dynamic filter ${\cal H}\left( s \right)\left[.\right] = {\textstyle{\alpha  \over {s + \alpha }}}\left[.\right]{\rm{,\;}}\alpha  > 0$ is applied to the left- and right-hand sides of equation \eqref{eq1}:
\begin{equation}\label{eq6}
{z_f}\left( t \right) = \varphi _f^{\top}\left( t \right)\theta  + {w_f}\left( t \right){\text{,}} 
\end{equation}	
where
\begin{displaymath}
\begin{gathered}
    {z_f}\left( t \right){\text{:}} = \mathcal{H}\left( s \right)\left[z\left( t \right)\right]{\text{,\;}}{\varphi _f}\left( t \right){\text{:}} = \mathcal{H}\left( s \right)\left[\varphi \left( t \right)\right]{\text{,}}\;\\{w_f}\left( t \right){\text{:}} = \mathcal{H}\left( s \right)\left[w\left( t \right)\right].
\end{gathered}
\end{displaymath}	

Subtracting \eqref{eq6} from \eqref{eq1}, it is obtained:
\begin{equation}\label{eq7}
\tilde{z}\left( t \right) = {\phi ^{\top}}\left( t \right)\Theta  + f\left( t \right){\rm{,}}
\end{equation}
\begin{displaymath}
\begin{array}{c}
\tilde{z}\left( t \right){\rm{:}} = z\left( t \right) - {z_f}\left( t \right){\rm{,\;}}\phi \left( t \right){\rm{:}} = {{\begin{bmatrix}
{{\varphi ^{\top}}\left( t \right)}&{\varphi _f^{\top}\left( t \right)}
\end{bmatrix}}^{\top}}{\rm{,}}\\
f\left( t \right){\rm{:}} = w\left( t \right) - {w_f}\left( t \right){\rm{,\;}}\Theta {\text{:}} = {\mathcal{D}}\theta = {\begin{bmatrix}
\theta \\
{ - \theta }
\end{bmatrix}}.
\end{array}    
\end{displaymath}
{and $\mathcal{D}=\begin{bmatrix} I_{n\times n}&-I_{n \times n}
\end{bmatrix}^{\top}\in\mathbb{R}^{2n\times n}$ is a duplication matrix of full column rank.}

In the next step equation \eqref{eq7} is extended via \eqref{eq4a}:
\begin{equation}\label{eq8}
\begin{array}{l}
\dot Y\left( t \right) = {\textstyle{1 \over T}}\left[ \phi \left( t \right) {\tilde{z}\left( t \right)}
 -\phi \left( {t - T} \right){\tilde{z}\left( {t - T} \right)}\right]{\rm{,\;}}\\
\dot \Phi \left( t \right) = {\textstyle{1 \over T}}\left[ {\phi \left( t \right){\phi ^{\top}}\left( t \right) - \phi \left( {t - T} \right){\phi ^{\top}}\left( {t - T} \right)} \right]{\rm{, }}\\
Y\left( {{t_0}} \right) = {0_{2n}}{\rm{,\;}}\Phi \left( {{t_0}} \right) = {0_{2n \times 2n}}.
\end{array}
\end{equation}

Owing to the implication
\begin{displaymath}
\begin{gathered}
x\left( t \right) = {\textstyle{1 \over T}}\int\limits_{{\rm{max}}\left\{ {{t_0}{\rm{,\;}}t - T} \right\}}^t {x\left( s \right)ds}  \\ \Updownarrow \\
\dot x\left( t \right) ={\textstyle{1 \over T}}\left[ {x\left( t \right) - x\left( {t - T} \right)} \right]{\rm{,\;}}x\left( {{t_0}} \right) = 0{\rm{,}} 
\end{gathered}
\end{displaymath}
the signals $Y\left( t \right)$ and $\Phi\left( t \right)$ meet the following relation:
\begin{equation}\label{eq9}
Y\left( t \right) = \Phi \left( t \right)\Theta  + W\left( t \right){\rm{,}}
\end{equation}
where
\begin{displaymath}
\dot W\left( t \right) = {\textstyle{1 \over T}}\left[ {\phi \left( t \right)f\left( t \right) - \phi \left( {t - T} \right)f\left( {t - T} \right)} \right]{\rm{,\;}}W\left( {{t_0}} \right) = {0_{2n}}.
\end{displaymath}

Disturbance term in \eqref{eq9} always admits the decomposition:
\begin{equation}\label{eq10}
W\left( t \right){\rm{:}} = {{\cal L}_1}{\cal L}_1^{\top}W\left( t \right){\rm{ + }}{{\cal L}_2}{\cal L}_2^{\top}W\left( t \right){\rm{,}}
\end{equation}
where ${{\cal L}_1} \in {\mathbb{R}^{2n \times 2m}}$ and ${{\cal L}_2} \in {\mathbb{R}^{2n \times \left( {2n - 2m} \right)}}$ such that:
\begin{displaymath}
\begin{array}{c}
{\cal L}_1^{\top}{{\cal L}_1} = {I_{2m \times 2m}}{\rm{,\;}}{\cal L}_2^{\top}{{\cal L}_2} = {I_{\left( {2n - 2m} \right) \times \left( {2n - 2m} \right)}}{\rm{, }}\\
{{\cal L}_1}{\cal L}_1^{\top} + {{\cal L}_2}{\cal L}_2^{\top} = {I_{2n \times 2n}}{\rm{,}}
\end{array}    
\end{displaymath}
and $2\left( {n - m} \right)$ is the number of elements of the vector $\phi \left( t \right)$, for which the \emph{independence} condition holds:
\begin{equation}\label{eq11}
\mathop {{\rm{lim}}}\limits_{T \to \infty } {\textstyle{1 \over T}}\int\limits_{{\rm{max}}\left\{ {{t_0}{\rm{,\;}}t - T} \right\}}^t {{\phi _i}\left( s \right)f\left( s \right)ds}  = 0,\;\forall t\ge t_{0}.   
\end{equation}

{For example, condition \eqref{eq11} is met for $\phi_{i}\left(t\right)=\linebreak={\rm{sin}}\left(\omega_{1}t+c_{1}\right)$ and $f\left(t\right)={\rm{sin}}\left(\omega_{2}t+c_{2}\right)$ iff $\omega_{1}\ne\omega_{2}$.}

{It should be specially noted that as $\phi_{i}\left(t\right)$ and $\phi_{n+i}\left(t\right)$ are dependent, then if \eqref{eq11} is met for $\phi_{i}\left(t\right)$, then \eqref{eq11} is also met for $\phi_{n+i}\left(t\right)$. Therefore condition \eqref{eq11} is written only in terms of initial regressor and perturbation.} 

Multiplication of \eqref{eq9} by ${\rm{adj}}\left\{ {\Phi \left( t \right)} \right\}$ and substitution of \eqref{eq10} yields:
\begin{equation}\label{eq12}
{\cal Y}\left( t \right) = \Delta \left( t \right)\Theta  + {{\cal W}_1}\left( t \right) + {{\cal W}_2}\left( t \right){\rm{,}}
\end{equation}
where
\begin{displaymath}
\begin{array}{c}
{\cal Y}\left( t \right) {\rm{:}} =  {\rm{adj}}\left\{ {\Phi \left( t \right)} \right\}Y\left( t \right){\rm{,\;}}\Delta \left( t \right) {\rm{:}} =  {\rm{det}}\left\{ {\Phi \left( t \right)} \right\}{\rm{,}}\\
{{\cal W}_1}\left( t \right) {\rm{:}} = {\rm{adj}}\left\{ {\Phi \left( t \right)} \right\}{{\cal L}_1}{\cal L}_1^{\top}W\left( t \right){\rm{,\;}}\\
{{\cal W}_2}\left( t \right) {\rm{:}} = {\rm{adj}}\left\{ {\Phi \left( t \right)} \right\}{{\cal L}_2}{\cal L}_2^{\top}W\left( t \right).
\end{array}    
\end{displaymath}

{Consequently, perturbation $\mathcal{W}\left(t\right)$ is decomposed into two parts with different properties. The second part can be made negligibly small via large width $T > 0$ of the sliding window, but not the first part, which causes biased parameter estimates. Aforementioned decomposition motivates to consider two cases. For the first one we show that, using results of \cite{b17}, in the absence of $\mathcal{W}_{1}\left(t\right)$, the goal \eqref{eq5} can be trivially achieved. In the second case, using BELS, we demonstrate that, if \eqref{eq11} holds for sufficiently large number of $\varphi_{i}\left(t\right)$, then $\mathcal{W}_{1}\left(t\right)$ is annihilated, and the result of the first case is retrieved.} 

\textbf{Case 1)} {$2\left(n - m\right) = 2n$}, {\it i.e.}, from the point of view of the harmonic analysis, the disturbance spectrum has no common frequencies with the regressor one, and therefore, it holds that ${{\cal W}_1}\left( t \right) \equiv 0$ and ${{\cal L}_2}{\cal L}_2^{\top} = {I_{2n}}$. So the estimation law to meet \eqref{eq5} is designed on the basis of \eqref{eq12} using the results of \cite{b17}:
\begin{equation}\label{eq13}
\begin{array}{l}
\hat \theta \left( t \right) = \hat \kappa \left( t \right){{\cal L}_0}{\cal Y}\left( t \right){\rm{, }}\\
\dot {\hat \kappa} \left( t \right) =  - \gamma \Delta \left( t \right)\left( {\Delta \left( t \right)\hat \kappa \left( t \right) - 1} \right) - \dot \Delta \left( t \right){{\hat \kappa }^2}\left( t \right){\rm{,\;}}\\
\dot \Delta \left( t \right) = {\rm{tr}}\left( {{\rm{adj}}\left\{ {\Phi \left( t \right)} \right\}\dot \Phi \left( t \right)} \right),\\
\hat \kappa \left( {{t_0}} \right) = {{\hat \kappa }_0}{\rm{,\;}}\Delta \left( {{t_0}} \right) = 0{\rm{,}}
\end{array}    
\end{equation}
where ${{\cal L}_0} = {\begin{bmatrix}
{{I_{n \times n}}}&{{0_{n \times n}}}
\end{bmatrix}}$ and $\gamma  > 0$.

The properties of \eqref{eq13} are described in:

\begin{thm}
{Suppose that $\varphi \left( t \right){\rm{,\;}}w\left( t \right)$ are bounded and assume that:}
\begin{enumerate}
    \item[\textbf{C1)}] {there exist (possibly not unique) $T \ge {T_f} > 0$ and $\overline \alpha \ge \underline \alpha   > 0$ such that for all $t \ge {T_f}$ it holds that}
    \begin{equation}\label{eq14}\hspace{-20pt}
        0 < \underline \alpha  {I_{2n}} \le {\textstyle{1 \over T}}\int\limits_{{\rm{max}}\left\{ {{t_0}{\rm{,\;}}t - T} \right\}}^t {\phi \left( s \right){\phi ^{\top}}\left( s \right)ds}  \le \overline \alpha  {I_{2n}}{\rm{,}}        
    \end{equation}
    \item[\textbf{C2)}] {the condition \eqref{eq11} holds for all $i = 1{\rm{,}} \ldots {\rm{,2}}n$},
    \item[\textbf{C3)}] {$\gamma  > 0$ is chosen so that there exists $\eta  > 0$ such that}
	$\gamma {\Delta ^3}\left( t \right) \!+\! \Delta \left( t \right)\dot \Delta \left( t \right)\hat \kappa \left( t \right) \!+\! \dot \Delta \left( t \right) \!\ge\! \eta \Delta \left( t \right) > 0\;\forall t \ge {T_f}.$
\end{enumerate}

Then the estimation law \eqref{eq13} ensures that \eqref{eq5} holds.
\end{thm}
\emph{Proof of Theorem 1 is postponed to Appendix.}

If \textbf{C2} is violated, then, using proof of Theorem 1, it is obvious that the estimation law \eqref{eq13} asymptotically provides only biased estimates. To overcome this drawback, the main idea of {BELS is exploited in the second case.}

\textbf{Case 2)} {$n \le 2\left(n - m\right) < 2n$, {\it i.e.}, the spectrum of sufficiently large number of the elements of the regressor has} no common frequencies with the disturbance spectrum. To obtain the unbiased parameter estimates for this case, following \cite{b22,b23}, the perturbation ${{\cal L}_1}{\cal L}_1^{\top}W\left( t \right)$ will be expressed from the regression equation \eqref{eq12} and subtracted from equation \eqref{eq9}.

{As duplication matrix ${\mathcal{D}}\in\mathbb{R}^{2n\times n}$ has full column rank, then according to Proposition 1 from \cite{b22}, for all $n \ge 2m \ge 2$ there exists an annihilator ${{\cal H}}\in {\mathbb{R}^{2n \times 2m}}$} of full column rank such that
\begin{equation}\label{eq15}
    {{\cal H}^{\top}}{\mathcal{D}}=0_{2m \times n} \Rightarrow {{\cal H}^{\top}}\Theta  = 0_{{2m}}.
\end{equation}

Considering \eqref{eq15}, the multiplication of \eqref{eq12} firstly by $\mathcal{H}^{\top}$ and then by ${\text{adj}}\left\{ {{\mathcal{H}^{\top}}{\text{adj}}\left\{ {\Phi \left( t \right)} \right\}{\mathcal{L}_1}} \right\}$ yields:
\begin{equation}\label{eq16}
\begin{gathered}
    \mathcal{N}\left( t \right) = \mathcal{M}\left( t \right)\mathcal{L}_1^{\top}W\left( t \right) +\quad\quad\quad\quad\quad\quad\quad\quad\\\quad\quad\quad\quad\quad\quad
    +{\text{adj}}\left\{ {{\mathcal{H}^{\top}}{\text{adj}}\left\{ {\Phi \left( t \right)} \right\}{\mathcal{L}_1}} \right\}{\mathcal{H}^{\top}}{\mathcal{W}_2}\left( t \right){\text{,}}
\end{gathered}
\end{equation}	
where
\begin{displaymath}
\begin{gathered}
  \mathcal{N}\left( t \right){\text{:}} = {\text{adj}}\left\{ {{\mathcal{H}^{\top}}{\text{adj}}\left\{ {\Phi \left( t \right)} \right\}{\mathcal{L}_1}} \right\}{\mathcal{H}^{\top}}\mathcal{Y}\left( t \right){\text{, }} \hfill \\
  \mathcal{M}\left( t \right){\text{:}} = {\text{det}}\left\{ {{\mathcal{H}^{\top}}{\text{adj}}\left\{ {\Phi \left( t \right)} \right\}{\mathcal{L}_1}} \right\}. \hfill \\ 
\end{gathered}
\end{displaymath}
 
Now we are in position to \emph{annihilate} a part of perturbation term in \eqref{eq9} via simple substitution. For that purpose, equation \eqref{eq9} is multiplied by $\mathcal{M}\left( t \right)$, and ${\mathcal{L}_1}\mathcal{N}\left( t \right)$ is subtracted from the obtained result to write:
\begin{equation}\label{eq17}
\begin{gathered}
\lambda \left( t \right) = \Omega \left( t \right)\Theta  + \\ + \biggl[ \mathcal{M}\left( t \right){\mathcal{L}_2} - 
{\mathcal{L}_1}{\text{adj}}\big\{ {{\mathcal{H}^{\top}}{\text{adj}}\big\{ {\Phi \left( t \right)} \big\}{\mathcal{L}_1}} \big\}\times\hfill\quad\quad\quad\quad\\\quad\quad\quad\quad\quad\quad\quad\quad\quad\times{\mathcal{H}^{\top}}{\text{adj}}\big\{ {\Phi \left( t \right)} \big\}{\mathcal{L}_2} \biggr]\mathcal{L}_2^{\top}W\left( t \right){\text{,}}
\end{gathered}
\end{equation}
where
\begin{displaymath}
\begin{gathered}
  \lambda \left( t \right){\text{:}} = \mathcal{M}\left( t \right)Y\left( t \right) - {\mathcal{L}_1}\mathcal{N}\left( t \right){\text{,}} \hfill \\
  \Omega \left( t \right){\text{:}} = \mathcal{M}\left( t \right)\Phi \left( t \right). \hfill \\ 
\end{gathered}
\end{displaymath}

To obtain the regression equation with a regressor, which derivative is directly measurable, we use the following simple filtration ($k > 0$):
\begin{equation}\label{eq18}
\begin{gathered}
\begin{gathered}
  {{\dot \Omega }_f}\left( t \right) =  - k{\Omega _f}\left( t \right) + k\Omega \left( t \right){\text{,\;}}{\Omega _f}\left( {{t_0}} \right) = {0_{2n \times 2n}}{\text{,}} \hfill \\
  {{\dot \lambda }_f}\left( t \right) =  - k{\lambda _f}\left( t \right) + k\lambda \left( t \right){\text{,\;}}{\lambda _f}\left( {{t_0}} \right) = {0_{2n}}{\text{,}} \hfill \\ 
\end{gathered}
\end{gathered}
\end{equation}

Then, to convert \eqref{eq17} into a set of separate scalar regression equations, we multiply ${\lambda _f}\left( t \right)$ by ${\text{adj}}\left\{ {{\Omega _f}\left(t\right)} \right\}$:
\begin{equation}\label{eq19}
\begin{gathered}
\Lambda \left( t \right) = \omega \left( t \right)\Theta  + d\left(t\right){\text{,}}    
\end{gathered}
\end{equation}
where
\begin{displaymath}
\begin{array}{l}
\Lambda \left( t \right){\rm{:}} = {\rm{adj}}\big\{ {{\Omega _f}\left(t\right)} \big\}{\lambda _f}\left( t \right){\rm{,\;}}\omega \left( t \right){\rm{:}} = {\rm{det}}\big\{ {{\Omega _f}\left(t\right)} \big\}{\rm{,}}\\
d\left(t\right){\rm{:}}={\text{adj}}\big\{ {{\Omega _f}\left(t\right)} \big\}\tfrac{k}{{s + k}}\Biggl[\biggl[ \mathcal{M}\left( t \right){\mathcal{L}_2} -\\ -
{\mathcal{L}_1}{\text{adj}}\big\{ {{\mathcal{H}^{\top}}{\text{adj}}\big\{ {\Phi \left( t \right)} \big\}{\mathcal{L}_1}} \big\}{\mathcal{H}^{\top}}{\text{adj}}\big\{ {\Phi \left( t \right)} \big\}{\mathcal{L}_2} \biggr]\mathcal{L}_2^{\top}W\left( t \right)\Biggr].
\end{array}    
\end{displaymath}

{Equation \eqref{eq19} is an analogue of equation \eqref{eq12}, but with annihilated perturbation term $\mathcal{W}_{1}\left(t\right)$.} The following estimation law is introduced on the basis of the such equation:
\begin{equation}\label{eq20}
\begin{array}{l}
\hat \theta \left( t \right) = \hat \kappa \left( t \right){{\cal L}_0}\Lambda \left( t \right){\rm{,}}\\
\dot {\hat \kappa} \left( t \right) =  - \gamma \omega \left( t \right)\left( {\omega \left( t \right)\hat \kappa \left( t \right) - 1} \right) - \dot \omega \left( t \right){{\hat \kappa }^2}\left( t \right){\rm{,}}\;\\
\dot \omega \left( t \right) = {\rm{tr}}\left( {{\rm{adj}}\left\{ {{\Omega _f}\left( t \right)} \right\}{{\dot \Omega }_f}\left( t \right)} \right){\rm{,\;}}\\
\hat \kappa \left( {{t_0}} \right) = {{\hat \kappa }_0}{\rm{,\;}}\omega \left( {{t_0}} \right) = 0{\rm{,}}
\end{array}    
\end{equation}
where $\gamma  > 0$ stands for an adaptive gain.

The conditions, under which the stated goal \eqref{eq5} is achieved when the law \eqref{eq20} is applied, are described in:

\begin{thm}
{Suppose that $\varphi \left( t \right){\rm{,\;}}w\left( t \right)$ are bounded and assume that:}
\begin{enumerate}
    \item[\textbf{C1)}] {there exist (possibly not unique) $T \ge {T_f} > 0$ and $\underline \alpha   \ge \overline \alpha   > 0$ such that for all $t \ge {T_f}$ the inequality \eqref{eq14} holds},
    \item[\textbf{C2)}] {the condition \eqref{eq11} holds for $2\left( {n - m} \right) \ge {n}$ elements of $\phi \left( t \right)$},
    \item[\textbf{C3)}] {the eliminators ${{\cal L}_1} \in {\mathbb{R}^{2n \times 2m}}{\rm{,\;}}{{\cal L}_2} \in {\mathbb{R}^{2n \times \left( {2n - 2m} \right)}}$ are exactly known and such that there exist (possibly not unique) $T \ge {T_f} > 0$ and $\underline \beta   \ge \overline \beta   > 0$ such that for all $t \ge {T_f}$ it holds that:}
    \begin{displaymath}\label{eq21}
        \hspace{-25pt} 0 \!<\! \underline \beta  \! \le\! \left| {{\rm{det}}\!\left\{\! {{{\cal H}^{\top}}{\rm{adj}}\left\{ {{\textstyle{1 \over T}}\!\!\!\!\!\int\limits_{{\rm{max}}\left\{ {{t_0}{\rm{,\;}}t - T} \right\}}^t {\!\!\!\!\!\!\!\!\!\!\!\phi \left( s \right){\phi ^{\top}}\left( s \right)ds} } \right\}{{\cal L}_1}} \right\}} \right| \le \overline \beta  {\rm{,}}        
    \end{displaymath}
    \item[\textbf{C4)}] {$\gamma  > 0$ is chosen so that there exists $\eta  > 0$ such that}
    \begin{displaymath}
        \gamma {\omega ^3}\left( t \right) + \omega \left( t \right)\dot \omega \left( t \right)\hat \kappa \left( t \right) + \dot \omega \left( t \right) \ge \eta \omega \left( t \right) > 0\;\forall t \ge {T_f}.
    \end{displaymath}
\end{enumerate}

{Then the estimation law \eqref{eq20} ensures that \eqref{eq5} holds.}
    
\end{thm}

\emph{Proof of theorem 2 is given in \cite{b24}}

The obtained estimation law \eqref{eq20}, unlike \eqref{eq13}, guarantees that the goal \eqref{eq5} can be achieved so long as {sufficiently large number of the elements} of the regressor $\varphi \left( t \right)$ satisfies the condition \eqref{eq11}. Requirement \textbf{C1} is the condition of identifiability of $\Theta $ in the perturbation-free case. Requirements \textbf{C2} and \textbf{C3} are the conditions of identifiability of the perturbation ${\cal L}_1^{\top}W\left( t \right)$, restriction \textbf{C4} is necessary to satisfy convergence $\hat \kappa \left( t \right) \to {\omega ^{ - 1}}\left( t \right)$ as $t \to \infty $.

The main difficulty of the law \eqref{eq20} implementation is the need to know the elimination matrices ${{\cal L}_1} \in {\mathbb{R}^{2n \times 2m}}{\rm{,\;}}\linebreak{{\cal L}_2} \in {\mathbb{R}^{2n \times \left( {2n - 2m} \right)}}$. However, using some {\it a priori} information about the parameterization \eqref{eq1}, it is always possible to construct afromentioned matrices if the condition \textbf{C2} is satisfied. For example, if the signals $z\left( t \right)$ and ${\varphi ^{\top}}\left( t \right)$ are obtained via parameterization of a linear dynamical system {with relative degree one} \cite{b1,b2} ($\Lambda \left( s \right)$ denotes a monic Hurwitz polynomial of order ${n_y}$):
\begin{equation}\label{eq22}
\begin{array}{c}
z\left( t \right) = {\textstyle{{{s^{{n_y}}}} \over {\Lambda \left( s \right)}}}\left[y\left( t \right)\right]{\rm{,}}\\
\varphi \left( t \right) = {{\begin{bmatrix}
{ - {\textstyle{{\lambda _{{n_y} - 1}^{\top}\left( s \right)} \over {\Lambda \left( s \right)}}}\left[y\left( t \right)\right]}&{{\textstyle{{\lambda _{{n_y} - 1}^{\top}\left( s \right)} \over {\Lambda \left( s \right)}}}\left[u\left( t \right)\right]}
\end{bmatrix}}^{\top}}{\rm{, }}\\
\lambda _{{n_y} - 1}^{\top}\left( s \right) = {\begin{bmatrix}
{{s^{{n_y} - 1}}}& \cdots &s&1
\end{bmatrix}}{\rm{,}}
\end{array}
\end{equation}
and the input signal $u\left( t \right)$ does not depend from the output one $y\left( t \right)$, then the matrices ${{\cal L}_1} \in {\mathbb{R}^{2n \times 2m}}{\rm{,\;}}$ \linebreak ${{\cal L}_2} \in {\mathbb{R}^{2n \times \left( {2n - 2m} \right)}}$ are defined as follows ($n = 2{n_y}$):
\begin{displaymath}\label{eq23}
{{\cal L}_1} = {\begin{bmatrix}
{{I_{{n_y} \times {n_y}}}}&{{0_{2{n_y} \times {n_y}}}}\\
{{0_{{n_y} \times {n_y}}}}&{{I_{{n_y} \times {n_y}}}}\\
{{0_{2{n_y} \times {n_y}}}}&{{0_{{n_y} \times {n_y}}}}
\end{bmatrix}}{\rm{,\;}}{{\cal L}_2} = {\begin{bmatrix}
{{0_{{n_y} \times {n_y}}}}&{{0_{2{n_y} \times {n_y}}}}\\
{{I_{{n_y} \times {n_y}}}}&{{0_{{n_y} \times {n_y}}}}\\
{{0_{2{n_y} \times {n_y}}}}&{{I_{{n_y} \times {n_y}}}}
\end{bmatrix}}.    
\end{displaymath}

The requirement that $u(t)$ is independent from $y(t)$ is not restrictive, and the proposed identification algorithm is applicable to the identification in a closed-loop -- in such case the input signal is interpreted as a reference one.

\textbf{Remark 1.} {\emph{It should be specially noted, that, in some simple cases, there exists a ``good choice'' of $T$, which ensures disturbance annihilation without $T \to \infty$. For example, if $\phi_{i}\left(t\right)=1$, $f\left(t\right)={\rm{sin}}\left({\omega}t\right)$ and $T = {\textstyle{{2\pi } \over \omega }}$ then 
\begin{displaymath}
    {\textstyle{1 \over T}}\int\limits_{{\rm{max}}\left\{ {{t_0}{\rm{,\;}}t - T} \right\}}^t {{\phi _i}\left( s \right)f\left( s \right)ds}  = 0,\;\forall t\ge T. 
\end{displaymath}}}

\section{Numerical experiments}
The following system has been considered as an example:
\begin{equation}\label{eq24}
\begin{array}{l}
\dot x\left( t \right) = {\begin{bmatrix}
{x\left( t \right)}&{u\left( t \right)}
\end{bmatrix}}{\begin{bmatrix}
{{\theta _1}}\\
{{\theta _2}}
\end{bmatrix}} + \delta \left( t \right){\rm{,\;}}x\left( {{t_0}} \right) = {x_0}{\rm{,}}\\
y\left( t \right) = x\left( t \right)+v\left( t \right){\rm{,}}
\end{array}    
\end{equation}
where ${\theta _1} < 0$.

The control signal $u\left( t \right)$ and disturbances $\delta \left( t \right){\rm{,\;}}v\left( t \right)$ were chosen as follows:
\begin{equation}\label{eq25}
\begin{gathered}
u\left( t \right) = 10{\rm{sin}}\left( {0{.}2\pi t} \right){\rm{,\;}}\delta \left( t \right) = 5{\rm{sin}}\left( {0{.}6\pi t + {\textstyle{\pi  \over 4}}} \right){\rm{,\;}}\\
v\left( t \right) = 0.7 + {\rm{sin}}\left( {24\pi t + {\textstyle{\pi  \over 8}}} \right).        
\end{gathered}
\end{equation}

In such case equation \eqref{eq1} was defined as:
\begin{equation}\label{eq26}
\begin{array}{c}
z\left( t \right){\rm{:}} = {\textstyle{s \over {s + {\alpha _0}}}}\left[y\left( t \right)\right]{\rm{,\;}}\\
{\varphi ^{\top}}\left( t \right){\rm{:}} = {\begin{bmatrix}
{{\textstyle{1 \over {s + {\alpha _0}}}}\left[y\left( t \right)\right]}&{{\textstyle{1 \over {s + {\alpha _0}}}}\left[u\left( t \right)\right]}
\end{bmatrix}}{\rm{,\;}}\\
w\left( t \right){\rm{:}}\!=\!{\textstyle{s \over {s + {\alpha _0}}}}\left[v\left( t \right)\right] + {\textstyle{1 \over {s + {\alpha _0}}}}\left[\delta \left( t \right)\right]\!-\!{\theta _1}{\textstyle{1 \over {s + {\alpha _0}}}}\left[v\left( t \right)\right].\! 
\end{array}   
\end{equation}

As the control signal $u\left( t \right)$ did not depend from the disturbances $\delta \left( t \right){\rm{,\;}}v\left( t \right)$, then the conditions \textbf{C2} and \textbf{C3} from Theorem 2 were satisfied, and the elimination and annihilator matrices were chosen as:
\begin{displaymath}
{{\cal H}^{\top}} = {\begin{bmatrix}
1&0&1&0\\
0&1&0&1
\end{bmatrix}}{\rm{,\;}}{{\cal L}_1} = {\begin{bmatrix}
1&0\\
0&0\\
0&1\\
0&0
\end{bmatrix}}.    
\end{displaymath}

The parameters of the system \eqref{eq24}, filters \eqref{eq6}, \eqref{eq8}, \eqref{eq18}, \eqref{eq26} and estimation law \eqref{eq20} were picked as:
\begin{equation}\label{eq27}
{\alpha _0} = \alpha  = k = 10{\rm{,\;}}T = 25{\rm{,\;}}\gamma  = {10^{112}}.
\end{equation}

The high value of $\gamma $ could be explained by the fact that $\omega \left( t \right) \in \left( {{{10}^{ - 57}}{\rm{,\;}}{{10}^{ - 55}}} \right)$ for all $t \ge 25$ .

For comparison purposes, the gradient descent law based on \eqref{eq3} was also implemented:
\begin{equation}\label{eq28}
\dot {\hat \theta} \left( t \right) =  - {\gamma _\Delta }\Delta \left( t \right)\left( {\Delta \left( t \right)\hat \theta \left( t \right) - {\cal Y}\left( t \right)} \right){\rm{,}}    
\end{equation}
as well as the one with the averaging, which was proposed in \cite{b17}:
\begin{equation}\label{eq29}
\begin{array}{l}
{{\dot {\hat \theta} }_i}\left( t \right) =  - \frac{1}{{t + {F_0}}}\left( {{{\hat \theta }_i}\left( t \right) - {\vartheta _i}\left( t \right)} \right){\rm{,}}\;\\
{\vartheta _i}\left( t \right) = \hat \kappa \left( t \right){{\cal Y}_i}\left( t \right){\rm{,}}\\
\dot {\hat \kappa} \left( t \right) =  - {\gamma _\kappa }\Delta \left( t \right)\left( {\Delta \left( t \right)\hat \kappa \left( t \right) - 1} \right) - \dot \Delta \left( t \right){{\hat \kappa }^2}\left( t \right){\rm{,}}\;\\
\dot \Delta \left( t \right) = {\rm{tr}}\left( {{\rm{adj}}\left\{ {\Phi \left( t \right)} \right\}\dot \Phi \left( t \right)} \right){\rm{,}}\;\\
{{\hat \theta }_i}\left( {{t_0}} \right) = {{\hat \theta }_{0i}}{\rm{,\;}}\hat \kappa \left( {{t_0}} \right) = {{\hat \kappa }_0}{\rm{,\;}}\Delta \left( {{t_0}} \right) = 0{\rm{,}}
\end{array}    
\end{equation}
where $\Delta \left( t \right)$ and ${\cal Y}\left( t \right)$ were obtained with the help of \eqref{eq2a} + \eqref{eq2b} with $l = 1$.

{To demonstrate the awareness of estimators to track the system parameters change, the unknown parameters were set as $\theta = {\begin{bmatrix}
{ - 1}\\
1
\end{bmatrix}}$ for $t \le 150$ and $\theta={\begin{bmatrix}
{ - 0.75}\\
{0.5}
\end{bmatrix}}$ for $t > 150$.

The parameters of the laws \eqref{eq20}, \eqref{eq28}, \eqref{eq29} were set as:
\begin{equation}\label{eq30}
{\gamma _\kappa } = {10^4}{\rm{,\;}}{\gamma _\Delta } = {10^2}{\rm{,\;}}{F_0} = 0.01.
\end{equation}

Figure 1 depicts the behavior of the system \eqref{eq24} output in both disturbance-free case and the one when the perturbation was defined as in \eqref{eq25}. It illustrates that perturbations \eqref{eq25} noticeably affected the system output.
\begin{figure}[!thpb]\label{figure4}
\begin{center}
\includegraphics[scale=0.5]{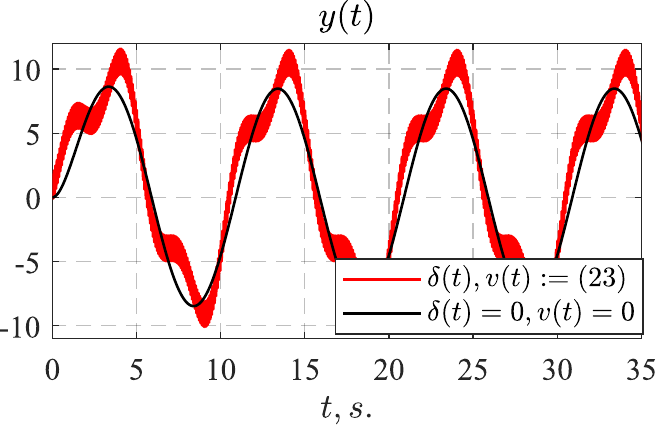}
\caption{{Behavior of $y\left( t \right)$ when $\delta \left( t \right){\rm{,\;}}v\left( t \right){\rm{:}} = 0$ and $\delta \left( t \right){\rm{,\;}}v\left( t \right){\rm{:}} = \eqref{eq25}$.}} 
\end{center}
\end{figure}

Figure 2a shows the behavior of the unknown parameter estimates when the laws \eqref{eq20}, \eqref{eq28}, \eqref{eq29} were applied. Figure 2b presents a comparison of $\hat \theta \left( t \right)$ transients for \eqref{eq20} using different values of the parameter $T$.

\begin{figure}[!thpb]
\begin{center}
\includegraphics[scale=0.55]{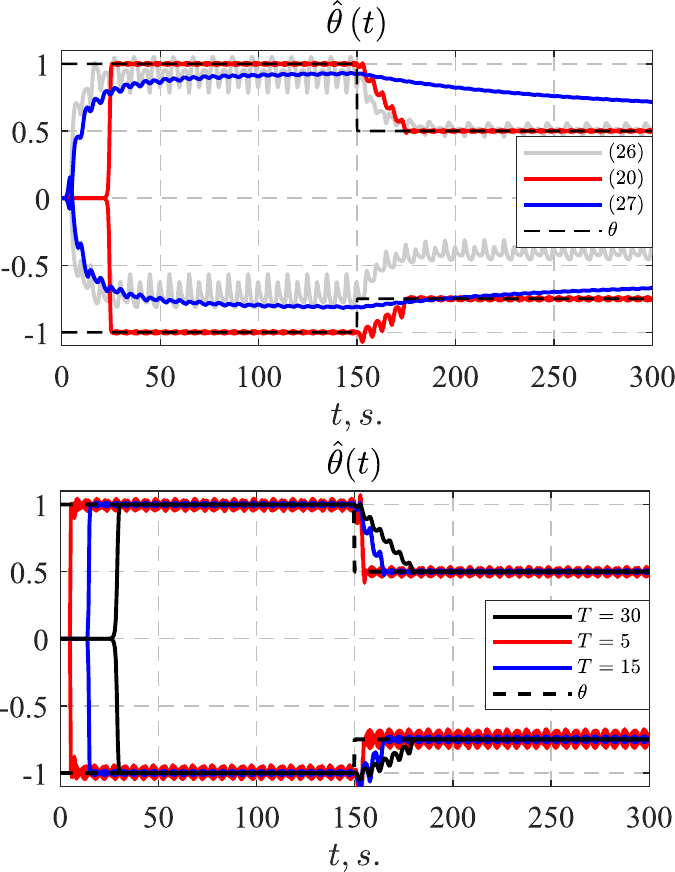}
\caption{{Behavior of a) $\hat \theta \left( t \right)$ for \eqref{eq20}, \eqref{eq28}, \eqref{eq29} and b) $\hat \theta \left( t \right)$ for \eqref{eq20} using different values of $T$.}} 
\end{center}
\end{figure}

The parametric error $\tilde \theta \left( t \right)$ for \eqref{eq28}, \eqref{eq29} remained bounded value and did not converge to zero even at $t \to \infty $. The proposed law \eqref{eq20} provided asymptotic convergence of the error $\tilde \theta \left( t \right)$ to an arbitrarily small neighborhood of zero defined by the parameter $T$.

\section{Conclusion}
Based on BELS approach, a modification of DREM procedure is proposed to ensure asymptotic convergence of the parametric error to an arbitrarily small neighborhood of zero defined by the arbitrary parameter $T$. To ensure convergence of the obtained estimates, independence of {sufficiently large number of  \emph{known} elements} of the regressor from the perturbation (\textbf{C2}) and the fulfillment of the conditions (\textbf{C1} and \textbf{C3}), similar to the well-known requirement of the regressor persistent excitation, are required.

The scopes of further research are to apply the proposed estimation law to the problems of design of adaptive observers and composite adaptive control systems and relax the conditions \textbf{C1} and \textbf{C3}.}

\bibliography{ifacconf}             
                                                   







\appendix
\section{Proof of Theorem 1}    
The regressor $\Delta \left( t \right)$ is defined as follows:
\begin{equation}\label{eqA1}
\Delta \left( t \right) = {\rm{det}}\left\{ {{\textstyle{1 \over T}}\int\limits_{{\rm{max}}\left\{ {{t_0}{\rm{,\;}}t - T} \right\}}^t {\phi \left( s \right){\phi ^{\top}}\left( s \right)ds} } \right\}{\rm{,}}    
\end{equation}

{Then, as \textbf{C1} is met, for all $t \ge {T_f}$ we have}
\begin{equation}\label{eqA3}
\Delta \left( t \right) \ge {\underline \alpha  ^{2n}} > 0
\end{equation}
and, consequently, the following error is well defined:
\begin{displaymath}
\tilde \kappa \left( t \right) = \hat \kappa \left( t \right) - {\Delta ^{ - 1}}\left( t \right){\rm{,}}    
\end{displaymath}
which is differentiated with respect to time and, owing to 
\begin{displaymath}
\begin{array}{c}
\Delta \left( t \right){\Delta ^{ - 1}}\left( t \right) = 1 \Leftrightarrow \dot \Delta \left( t \right){\Delta ^{ - 1}}\left( t \right) + \Delta \left( t \right){\textstyle{{d{\Delta ^{ - 1}}\left( t \right)} \over {dt}}} = 0{\rm{,}}\\
 \Updownarrow \\
{\textstyle{{d{\Delta ^{ - 1}}\left( t \right)} \over {dt}}} =  - \dot \Delta \left( t \right){\Delta ^{ - 2}}\left( t \right){\rm{,}}
\end{array}    
\end{displaymath}	
it is obtained:
\begin{equation}\label{eqA4}
\begin{array}{l}
\dot {\tilde \kappa}  =  - \gamma \Delta \left( {\Delta \hat \kappa  - 1} \right) - \dot \Delta {{\hat \kappa }^2} + \dot \Delta {\Delta ^{ - 2}} = \\
 =  - \gamma {\Delta ^2}\tilde \kappa  - \dot \Delta \left( {\hat \kappa  + {\Delta ^{ - 1}}} \right)\tilde \kappa  =  \\
 = - \left( {\gamma {\Delta ^2} + \dot \Delta \hat \kappa  + \dot \Delta {\Delta ^{ - 1}}} \right)\tilde \kappa {\rm{,}}
\end{array}    
\end{equation}
where $\dot \Delta \left( t \right)$ obeys Jacobi’s formula:
\begin{displaymath}
\dot \Delta \left( t \right) = {\rm{tr}}\left( {{\rm{adj}}\left\{ {\Phi \left( t \right)} \right\}\dot \Phi \left( t \right)} \right){\rm{,\;}}\Delta \left( {{t_0}} \right) = 0.    
\end{displaymath}

The quadratic form $V\left( t \right) = {\textstyle{1 \over 2}}{\tilde \kappa ^2}\left( t \right)$ is introduced, which derivative is written as:
\begin{displaymath}
\dot V\left( t \right) =  - 2\left( {\gamma {\Delta ^2}\left( t \right) + \dot \Delta \left( t \right)\hat \kappa \left( t \right) + \dot \Delta \left( t \right){\Delta ^{ - 1}}\left( t \right)} \right)V\left( t \right){\rm{,}}   
\end{displaymath}
from which, when $\gamma {\Delta ^3}\left( t \right) + \Delta \left( t \right)\dot \Delta \left( t \right)\hat \kappa \left( t \right) + \dot \Delta \left( t \right) \ge \eta \Delta \left( t \right) > 0\;\forall t \ge {T_f}$, then for all $t \ge {T_f}$ there exists the following upper bound:
\begin{equation}\label{eqA5}
\left| {\tilde \kappa \left( t \right)} \right| \le {e^{ - \eta \left( {t - {T_f}} \right)}}\left| {\tilde \kappa \left( {{t_0}} \right)} \right|.    
\end{equation}

For all $t \ge {T_f}\;$  $\hat \theta \left( t \right)$ is rewritten in the following form:
\begin{equation}\label{eqA6}
\begin{array}{l}
\hat \theta \left( t \right) = \hat \kappa \left( t \right){{\cal L}_0}{\cal Y}\left( t \right) \pm {\Delta ^{ - 1}}\left( t \right){{\cal L}_0}{\cal Y}\left( t \right) = \\
= {\Delta ^{ - 1}}\left( t \right){{\cal L}_0}{\cal Y}\left( t \right) + \tilde \kappa \left( t \right){{\cal L}_0}{\cal Y}\left( t \right) = \\
 = \theta  + {\Delta ^{ - 1}}\left( t \right){{\cal L}_0}{{\cal W}_2}\left( t \right) + \tilde \kappa \left( t \right){{\cal L}_0}{\cal Y}\left( t \right) = \\
 = \theta  + {\Delta ^{ - 1}}\left( t \right){{\cal L}_0}{\rm{adj}}\left\{ {\Phi \left( t \right)} \right\}{{\cal L}_2}{\cal L}_2^{\top}W\left( t \right) + \tilde \kappa \left( t \right){{\cal L}_0}{\cal Y}\left( t \right) = \\
 = \theta  + {{\cal L}_0}{\Phi ^{ - 1}}\left( t \right)W\left( t \right) + \tilde \kappa \left( t \right){{\cal L}_0}{\cal Y}\left( t \right).
\end{array}
\end{equation}

When \textbf{C1} is met, for all $t \ge {T_f}$ it holds that
\begin{equation}\label{eqA7}
{\Phi ^{ - 1}}\left( t \right) = {\left[ {{\textstyle{1 \over T}}\int\limits_{t - T}^t {\phi \left( s \right){\phi ^{\top}}\left( s \right)ds} } \right]^{ - 1}} \ge {\overline \alpha  ^{ - 1}}{I_{2n}}{\rm{,}}  
\end{equation}
and consequently, from \eqref{eqA6} we have the following upper bound of the error $\tilde \theta \left( t \right)$:
\begin{displaymath}\label{eqA8}
\begin{array}{l}
\left\| {\tilde \theta \left( t \right)} \right\| \!\le\! \left\| {{\Phi ^{ - 1}}\left( t \right)} \right\|\left\| {{\textstyle{1 \over T}}\int\limits_{t - T}^t {\phi \left( s \right)f\left( s \right)ds} } \right\| \!+\!\left| {\tilde \kappa \left( t \right)} \right|\left\| {{\cal Y}\left( t \right)} \right\| \\
 \le {\underline \alpha  ^{ - 1}}\left\| {{\textstyle{1 \over T}}\int\limits_{t - T}^t {\phi \left( s \right)f\left( s \right)ds} } \right\| + {e^{ - \eta \left( {t - {T_f}} \right)}}\left| {\tilde \kappa \left( {{t_0}} \right)} \right|\left\| {{\cal Y}\left( t \right)} \right\|{\rm{,}}
\end{array}    
\end{displaymath}
from which, as $\left\| {{\cal Y}\left( t \right)} \right\|$ is bounded, for bounded $\varphi \left( t \right){\rm{,\;}}w\left( t \right)$ it is obtained:
\begin{displaymath}\label{eqA9}
\begin{array}{l}
\mathop {{\rm{lim}}}\limits_{t \to \infty } \left\| {\tilde \theta \left( t \right)} \right\| \le {\underline \alpha  ^{ - 1}}\left\| {\mathop {{\rm{lim}}}\limits_{t \to \infty } {\textstyle{1 \over T}}\int\limits_{t - T}^t {\phi \left( s \right)f\left( s \right)ds} } \right\|{\rm{:}} = \varepsilon \left( T \right){\rm{,}}\\
\mathop {{\rm{lim}}}\limits_{T \to \infty } \varepsilon \left( T \right) = {\underline \alpha  ^{ - 1}}\left\| {\mathop {{\rm{lim}}}\limits_{t \to \infty } \mathop {{\rm{lim}}}\limits_{T \to \infty } {\textstyle{1 \over T}}\int\limits_{t - T}^t {\phi \left( s \right)f\left( s \right)ds} } \right\| = 0{\rm{,}}
\end{array}    
\end{displaymath}
which was to be proved.
\end{document}